\documentclass[letterpaper]{IEEEtran}         
\usepackage{amsmath,graphicx}
\usepackage{amsfonts}
\usepackage{enumerate}
\usepackage{bm}
\usepackage{algorithm} 
\usepackage{algorithmic} 

\usepackage{mathrsfs}
\usepackage{stfloats}
\usepackage{enumerate}
\usepackage{epstopdf}
\usepackage{subfigure}
\usepackage{float}
\usepackage{amsbsy}
\usepackage{amsmath}
\usepackage{amssymb}
\usepackage{color}
\usepackage{ntheorem}
\usepackage{cite}
\theorembodyfont{\upshape}
\theoremheaderfont{\rmfamily\itshape}
\theoremseparator{:}
\theoremstyle{remark}

\newtheorem{theo}{\hspace{1em}Theorem}
\newtheorem{lemma}{\hspace{1em}Corollary}
\newtheorem*{pproof}{\hspace{2em}Proof}

\begin{document}
\title{\huge{Near-Field Modelling and Performance Analysis of Modular Extremely Large-Scale Array Communications}\vspace{-1pt}}
\author{\IEEEauthorblockN{Xinrui~Li, Haiquan~Lu, Yong Zeng,  \emph{Member, IEEE}, Shi Jin, \emph{Senior Member, IEEE}, and Rui Zhang, \emph{Fellow, IEEE}
\thanks{This work was supported by the National Key R$\&$D Program of China with
Grant number 2019YFB1803400.}
\thanks{X. Li, H. Lu, Y. Zeng, and S. Jin are with the National Mobile Communications Research
Laboratory, Southeast University, Nanjing 210096, China. Y. Zeng is also
with the Purple Mountain Laboratories, Nanjing 211111, China (e-mail: \{230218659, haiquanlu, yong\_zeng, jinshi\}@seu.edu.cn). (\emph{Corresponding author: Yong Zeng}.)}
\thanks{R. Zhang is with the Department of Electrical and Computer Engineering, National University of Singapore, Singapore 117583 (e-mail: elezhang@nus.edu.sg).}
}}
\maketitle
\vspace{-1cm}

\begin{abstract}
 This letter studies a new array architecture,
 termed as \textit{modular extremely large-scale array (XL-array)},
 for which a large number of array elements are arranged in a modular manner.
 Each module consists of a moderate number of array elements
 and the modules are regularly arranged with the inter-module space typically much larger than signal wavelength
 to cater to the actual mounting structure.
 We study the mathematical modelling and conduct the performance analysis for modular XL-array communications,
 by considering the non-uniform spherical wave (NUSW) characteristic that is more suitable than
 the conventional uniform plane wave (UPW) assumption for physically large arrays.
  A closed-form expression is derived for the maximum signal-to-noise ratio (SNR) in terms of the geometries of the modular XL-array,
  including the total array size and module separation, as well as the user's location.
  The asymptotic SNR scaling law is revealed as the size of modular array goes to infinity.
  Furthermore, we show that the developed modelling and performance analysis include the existing results for collocated XL-array
  or far-field UPW assumption as special cases.
  Numerical results demonstrate the importance of near-field modelling for modular XL-array communications
  since it leads to significantly different results from the conventional far-field UPW modelling.
\end{abstract}



\IEEEpeerreviewmaketitle
\section{Introduction}
With the commercial deployment of the fifth-generation (5G) mobile communication networks,
massive multiple-input multiple-output (MIMO), as a key enabling technology for spectrum-efficient communication,
has become a reality \cite{Zhang2020, Bjornson2019}.
To support the ambitious goals of 6G wireless networks, such as ultra-high
throughput, ultra-low latency, and ultra-high reliability \cite{Bii2019},
there has been growing interest in further increasing the antenna number/size beyond the current massive MIMO systems,
known as extremely large-scale MIMO (XL-MIMO) \cite{Lu2021, Zeng2021},
extremely large aperture arrays (ELAAs) \cite{Bjornson2019} or ultra-massive MIMO (UM-MIMO) \cite{Akyildiz2016}.\par
As the number of antennas keeps increasing,
there are in general two existing architectures to accommodate the physically and electrically large antennas,
namely \textit{collocated extremely large-scale array (XL-array)} \cite{Lu2021, Zeng2021} and \textit{distributed antenna system (DAS)} \cite{Choi2020}.
Specifically, for collocated XL-array architecture,
all antenna elements are arranged regularly in the same platform with adjacent elements separated in wavelength scale,
so as to form the standard antenna arrays.
However, the size of collocated XL-array are practically limited by the available contiguous space of the mounting structure \cite{Bjornson2019}.
On the other hand, for DAS, the antennas are scattered in geographically separated locations,
which are inter-connected by high-capacity and low-latency backhaul/fronthaul links to support joint signal processing \cite{Choi2020}.
Some representative examples
for DAS architecture include network MIMO \cite{kara2006}, the cloud radio
access network (C-RAN) \cite{Yoon2015}, and the emerging cell-free
massive MIMO  \cite{Ngo2017}.
However, DAS usually requires not only a large number of sites for antenna deployment,
but also the prohibitive backhaul/fronthaul capacity
and sophisticated coordination among different sites \cite{Ngo2017}.\par
To mitigate the aforementioned drawbacks of collocated XL-array and DAS architectures,
in this letter, we investigate a new array architecture,
termed \textit{modular XL-array},
for which the array elements are arranged in a modular manner, as illustrated in Fig. 1.
Each module consists of a moderate number of array elements,
and different modules are arranged regularly on a common platform with inter-module spacing typically much larger than signal wavelength to cater to the practical mounting structure.
Compared to the collocated XL-array architecture,
modular XL-array is much more flexible for deployment, thanks to its conformal capability with its mounting structure.
On the other hand, in contrast to the DAS architecture,
modular XL-array can achieve large-scale joint signal processing without having to perform sophisticated
inter-site signal exchange or coordination.\par
It is worth mentioning that as the size of antenna array goes extremely large,
the conventional uniform plane wave (UPW) assumption may no longer hold \cite{Lu2021}.
Instead, the more general non-uniform spherical wave (NUSW) characteristics should be taken into account to more accurately model the variations of signal phase and amplitude across array elements,
and some preliminary works along this direction have been carried out in \cite{Zeng2021,Hu2021,Dar2020}.
However, these existing works mainly focus on
the collocated XL-array architecture, while their results cannot be
directly applied to the new modular XL-array architecture.\par
\begin{figure}[t]
\begin{centering}
\includegraphics[scale=0.62]{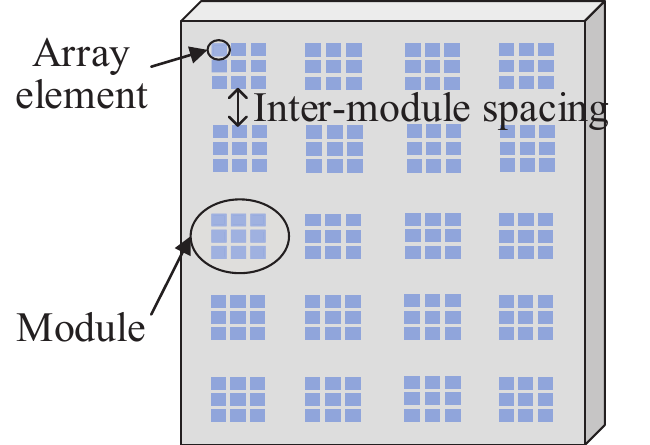}
\vspace{-0.3cm}
\caption{An illustration of modular XL-array mounted on building facades.} \label{picture0}
\end{centering}
\vspace{-0.7cm}
\end{figure}

To fill the above gap,
this letter pursues the mathematical modelling and performance analysis for modular XL-array communications.
By considering the NUSW characteristics,
a closed-form expression of the maximum signal-to-noise ratio (SNR) with the optimal maximal-ratio combining (MRC) beamforming for the basic modular extremely large-scale uniform linear array (XL-ULA) is derived.
The result shows that the maximum SNR for modular XL-ULA depends on both the array geometry, including its total array size and module separation, and the user's location.
The asymptotic analysis shows that as the array size increases by adding more modules,
the SNR approaches to a constant value that depends on the module separation and the user's projected distance to the modular XL-array.
Furthermore, we show that the obtained SNR expression includes the existing results for collocated XL-array or far-field UPW assumption as special cases.
Numerical results demonstrate the importance of near-field modelling for modular XL-array communications.\par

\section{System Model} \label{model}\vspace{-1pt}

\begin{figure}[t]
\begin{centering}
\includegraphics[scale=0.7]{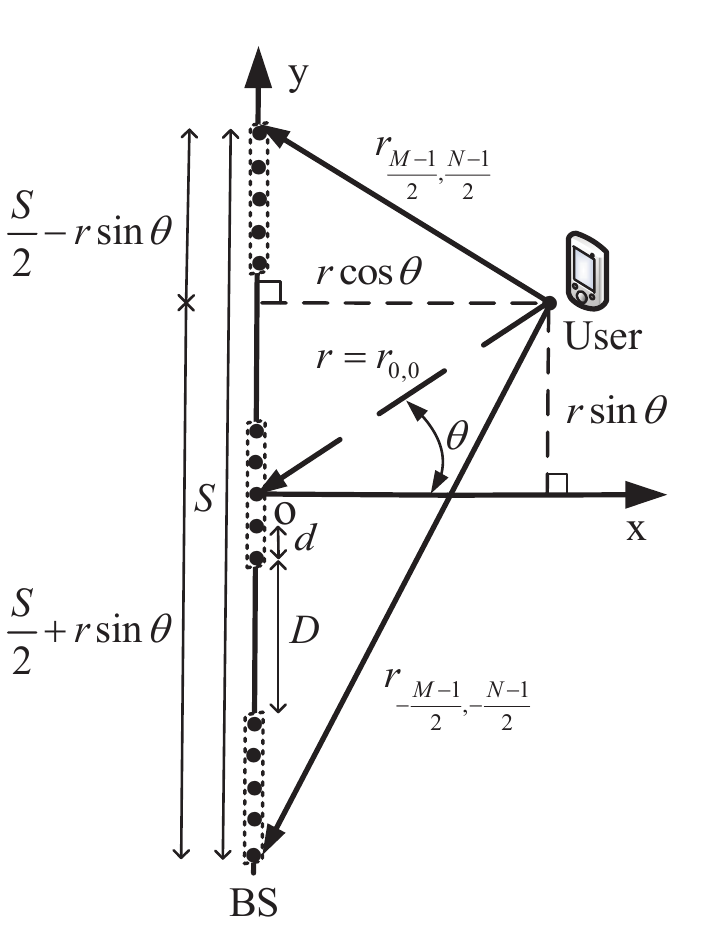}
\vspace{-0.3cm}
\caption{A wireless communication system with modular XL-ULA.} \label{picture1}
\end{centering}
\vspace{-0.6cm}
\end{figure}

As shown in Fig. 2, we consider a wireless communication system in the near-field region,
where a base station (BS) equipped with a modular XL-array communicates with a single-antenna user.
The total number of modules of the modular XL-array is denoted as $N$,
and each module consists of $M$ antenna elements, with inter-element spacing denoted by $d$.
Therefore, the total number of array elements is $MN$.
Furthermore, denote by $D$ the inter-module separation.
Typically, the inter-element spacing $d$ is on the wavelength scale (e.g., $d=\lambda/2$),
where $\lambda$ denotes the signal wavelength,
while the inter-module spacing $D$ is much larger than the wavelength,
so as to cater to the actual mounting structure.
For example, modular XL-array can be integrated into facades of buildings in airports, shopping malls, etc.,
where different modules are separated by windows \cite{Bjornson2019}.
For notational convenience, let $D=Ld$, for some integer $L\geq 1$.
It is not difficult to see that when $D=d$ or $L=1$,
the modular XL-array degenerates to the conventional collocated XL-array \cite{Lu2021}.\par
For convenience, we assume that both $N$ and $M$ are odd number,
and the modular XL-array is placed along the $\emph{y}$-axis
that is symmetric around the origin.
Thus, the position of the $m$th array element of module $n$,
where $n= 0,\pm 1,...,\pm \frac{N-1}{2}$, $m= 0,\pm 1,...,\pm \frac{M-1}{2}$, can be written as ${\bf w}_{m,n}=[0,y_{m,n}]^T$,
where $y_{m,n}=n[D+(M-1)]d+md=(Kn+m)d$, with $K=M+L-1$.
Therefore, the physical dimension of the modular XL-ULA can be expressed as $S=[K(N-1)+(M-1)]d$.\par
Furthermore, let $r$ denote the distance of the user from the center of the modular XL-array, and
$\theta \in [-\frac{\pi}{2},\frac{\pi}{2}]$ denote the user's direction with respect to the positive $\emph{x}$-axis.
As a result, the coordinate of the user's location is ${\bf q}=[r\cos{\theta},r\sin{\theta}]^T$.
The distance between the user and the $m$th antenna element of module $n$ is then given by
\begin{equation}\label{EQU-1} \vspace{-3pt}
\begin{split}
r_{m,n}&=||{\bf q}-{\bf w}_{m,n}||\\
&=r\sqrt{1-2(Kn+m)\epsilon\sin{\theta}+(Kn+m)^2\epsilon^2},\\
\end{split}
\end{equation}
where $\epsilon \triangleq \frac{d}{r}$. Note that since inter-element separation $d$ is at the order of signal wavelength,
we have $\epsilon \ll 1$.\par

Considering the free-space line-of-sight (LoS) propagation\footnote{The investigation of modular XL-array communication for non-line-of-sight (NLoS) propagation is left as the future work.},
 the array response vector ${\bf a}(r,\theta) \in \mathbb{C}^{(MN)\times 1}$
between the user and the modular XL-ULA can be expressed as
\begin{equation}\label{EQU-21} \vspace{-3pt}
\begin{split}
{\bf a}(r,\theta)=[{\bf a}^T_{-\frac{N-1}{2}}(r,\theta),...,{\bf a}^T_0(r,\theta),...,{\bf a}^T_{\frac{N-1}{2}}(r,\theta)]^T,\\
\end{split}
\end{equation}
where ${\bf a}_n(r,\theta) \in \mathbb{C}^{M\times 1}$ is the array response vector for module $n$, given by
\begin{equation}\label{EQU-2} \vspace{-3pt}
{\bf a}_n(r,\theta)=[a_{-\frac{M-1}{2},n}(r,\theta),...,a_{0,n}(r,\theta),...,a_{\frac{M-1}{2},n}(r,\theta)]^T,
\end{equation}
where $a_{m,n}(r,\theta)=\frac{\sqrt{\beta_0}}{r_{m,n}}e^{-j\frac{2\pi}{\lambda}r_{m,n}}$ with $\beta_0$
denoting the channel power at the reference distance $d_0=1$ $\rm m$.\par
Note that different from the conventional far-field modelling with UPW assumption, i.e.,
$a_{m,n}(r,\theta)=\frac{\sqrt{\beta_0}}{r}e^{-j\frac{2\pi}{\lambda}[r-(Kn+m)d\sin\theta]}$,
which only depends on the signal angle,
$a_{m,n}(r,\theta)$ in \eqref{EQU-21} for the near-field NUSW modelling depends both on both amplitude and phase
variations.\par
We focus on the uplink communication,
while the results can be straightforwardly extended to the downlink scenario.
The resulting signal at the BS after receive beamforming is
\begin{equation}\label{EQU-3} \vspace{-3pt}
{y}={\bf v}^H{\bf a}(r,\theta)\sqrt{P}s+{\bf v}^H{\bf z},
\end{equation}
where ${\bf v}\in\mathbb{C}^{(MN)\times 1}$ is the receive beamforming vector with $||{\bf v}||=1$, $P$ is the transmit power,
$s$ is the information-bearing signal and ${\bf z}$ denotes the additive white Gaussian noise (AWGN), following
the distribution of a circular symmetric complex Gaussian vector with mean vector $\bf 0$ and covariance matrix ${\sigma}^2 {\bf I}_{MN}$, denoted as
${\bf z}\sim{\cal CN}({\bf 0},{\sigma}^2 {\bf I}_{MN})$.\par

Therefore, the resulting SNR can be written as
\begin{equation}\label{EQU-4} \vspace{-3pt}
\gamma_{\rm NUSW}={\overline P}|{\bf v}^H {{\bf a}(r,\theta)}|^2,
\end{equation}
where ${\overline P}=\frac{P}{\sigma^2}$ is the transmit SNR.
It is well known that for single-user communication,
the MRC beamforming is optimal, which is given by ${\bf v}^*=\frac{{\bf a}(r,\theta)}{||{{\bf a}(r,\theta)}||}$,
and the resulting maximum SNR is $\gamma_{\rm NUSW}={\overline P}||{\bf a}(r,\theta)||^2$.
After substituting ${\bf a}(r,\theta)$ using (1)-(3), 
the maximum SNR can be written as \eqref{EQU-5}, shown at the top of the next page.\par
\newcounter{TempEqCnt} 
\setcounter{TempEqCnt}{\value{equation}} 
\setcounter{equation}{5} 
\begin{figure*}[ht] 
	\begin{equation}\label{EQU-5}
	\gamma_{\rm NUSW}={{\overline P}||{{\bf a}(r,\theta)}||^2}=\frac{\overline P \beta_0}{r^2}\sum_{n=-\frac{N-1}{2}}^{\frac{N-1}{2}}\sum_{m=-\frac{M-1}{2}}^{\frac{M-1}{2}}\frac{1}{1-2m\epsilon\sin\theta -2K n\epsilon\sin\theta +2Kmn\epsilon^2+m^2\epsilon^2+K^2n^2\epsilon^2}.\\
	\end{equation}
{\noindent} \rule[-10pt]{17.5cm}{0.05em}
\end{figure*}
\section{Closed-Form Expression and Performance Analysis} \label{model}\vspace{-1pt}
In this section,
we first derive the closed-form expression for the maximum SNR in \eqref{EQU-5},
and then demonstrate that the obtained new results
include the near-field collocated XL-array and far-field UPW models as special cases.
Furthermore, asymptotic performance analysis is provided to reveal the near-field SNR scaling law
when the size of the modular XL-array goes asymptotically large.
\begin{theo}\label{theo1}
For modular XL-array communication,
the maximum SNR can be approximated in closed-form as

\begin{equation}\label{EQU-6} \vspace{-3pt}
\begin{split}
&\gamma_{\rm NUSW}\approx\frac{{\overline P}\beta_0}{(M-1)d^2+Dd}\left[h \left(\frac{S_1}{2r\cos\theta}-\tan\theta \right)\right.\\
&+h \left(\frac{S_1}{2r\cos\theta}+\tan\theta \right)-h \left(\frac{S_1-2Md}{2r\cos\theta}-\tan\theta \right)\\
&\left.-h \left(\frac{S_1-2Md}{2r\cos\theta}+\tan\theta \right) \right],\\
\end{split}
\end{equation}
where  $h(x)\triangleq x\arctan x-\frac{1}{2}\ln(1+x^2)$ and $S_1\triangleq S+Md+D$.
\end{theo}
\begin{pproof}Please refer to Appendix A. $\hfill \blacksquare$\end{pproof}  \par
Theorem \ref{theo1} shows that the maximum SNR for modular XL-ULA communication is governed by the geometries of the modular XL-array,
such as the total array size $S$ and module separation $D$,
as well as the user's location $\bf q$. Specifically, when $S_1 \approx S$ holds at the case of $N\gg 1$ with relatively small $M$ and $D$,
we have $\frac{S_1}{2r\cos\theta}-\tan\theta  \approx \frac{\frac{S}{2}-r\sin\theta}{r\cos\theta} $
and $\frac{S_1}{2r\cos\theta}+\tan\theta \approx \frac{\frac{S}{2}+r\sin\theta}{r\cos\theta}$, and detailed geometric relations are shown in Fig. 2. To obtain more insight,
we first consider the special case when $D=d$ or $L=1$,
under which the modular XL-array reduces to the conventional collocated XL-array \cite{Lu2021}.\par
\begin{lemma}\label{lemma1}
When $D=d$ or $L=1$, $\gamma_{\rm NUSW}$ in \eqref{EQU-6} reduces to
\begin{equation}\label{EQU-7151} \vspace{-3pt}
\begin{split}
\gamma_{\rm NUSW}&\approx\frac{{\overline P}\beta_0}{r d \cos\theta}\left[\arctan(\frac{MNd}{2r\cos\theta}-\tan\theta)\right.\\
&\left.+\arctan(\frac{MNd}{2r\cos\theta}+\tan\theta)\right].\\
\end{split}
\end{equation}
\end{lemma}\par
\begin{pproof} Please refer to Appendix B.$\hfill \blacksquare$ \end{pproof}\par
It is noted that compared with the result in \eqref{EQU-6}, determined by the geometries of the modular XL-array,
the obtained SNR in \eqref{EQU-7151} only scales with the total array number/size $MN$,
which is consistent with the SNR for the conventional collocated XL-array in \cite{Lu2021}. \par

\begin{lemma}\label{lemma2}
For modular XL-ULA communication,
the asymptotical SNR with infinitely large array by increasing the module number $N$ is
\begin{equation}\label{EQU-7} \vspace{-3pt}
\lim_{N \to \infty} \gamma_{\rm NUSW}=\frac{\pi M\overline{P}\beta_0 }{[(M-1)d+D] r\cos\theta}.
\end{equation}
\end{lemma}
\begin{pproof} Please refer to Appendix C.$\hfill \blacksquare$ \end{pproof}\par
It is observed from \eqref{EQU-7} that the obtained SNR approaches to a constant value as $N$ goes to infinity.
Besides, in contrast to the asymptotic SNR for the infinitely large-scale collocated array that only depends on the user's projected distance to the array, i.e., $r\cos \theta$ \cite{Lu2021},
the obtained SNR in \eqref{EQU-7} not only is inversely proportional to $r\cos\theta$, but also decreases as module separation $D$ increases. This is expected since $D$ in practice increases the distance between parts of array elements and the user, and hence decreases the SNR.\par
Moreover, when the link distance is much larger than $S_1$, 
the SNR expression degenerates to the result of the conventional UPW model,
which is shown in the following Corollary.\par
\begin{lemma}\label{lemma3}
When $r \gg \frac{1}{2}S_1$, $\gamma_{\rm NUSW}$ in \eqref{EQU-6} reduces to
\begin{equation}\label{EQU-11} \vspace{-3pt}
\gamma_{\rm NUSW}\approx \gamma_{\rm UPW}= \frac{MN\overline{P}\beta_0}{r^2}.
\end{equation}
\end{lemma}
\begin{pproof} Please refer to Appendix D.$\hfill \blacksquare$ \end{pproof}\par
Corollary \ref{lemma3} shows that the new closed-form SNR generalizes the result based on the conventional UPW,
and it is applicable for both near-field and far-field scenarios.
It is also observed that as the array number goes to infinity, $\gamma_{\rm UPW}$ with the conventional UPW modelling grows unbounded, which is impractical for modular XL-array communication in the near-field region.\par

\section{Numerical results}\label{result} \vspace{-1pt}
\begin{figure}[t]
  \begin{centering}
  \includegraphics[scale=0.42]{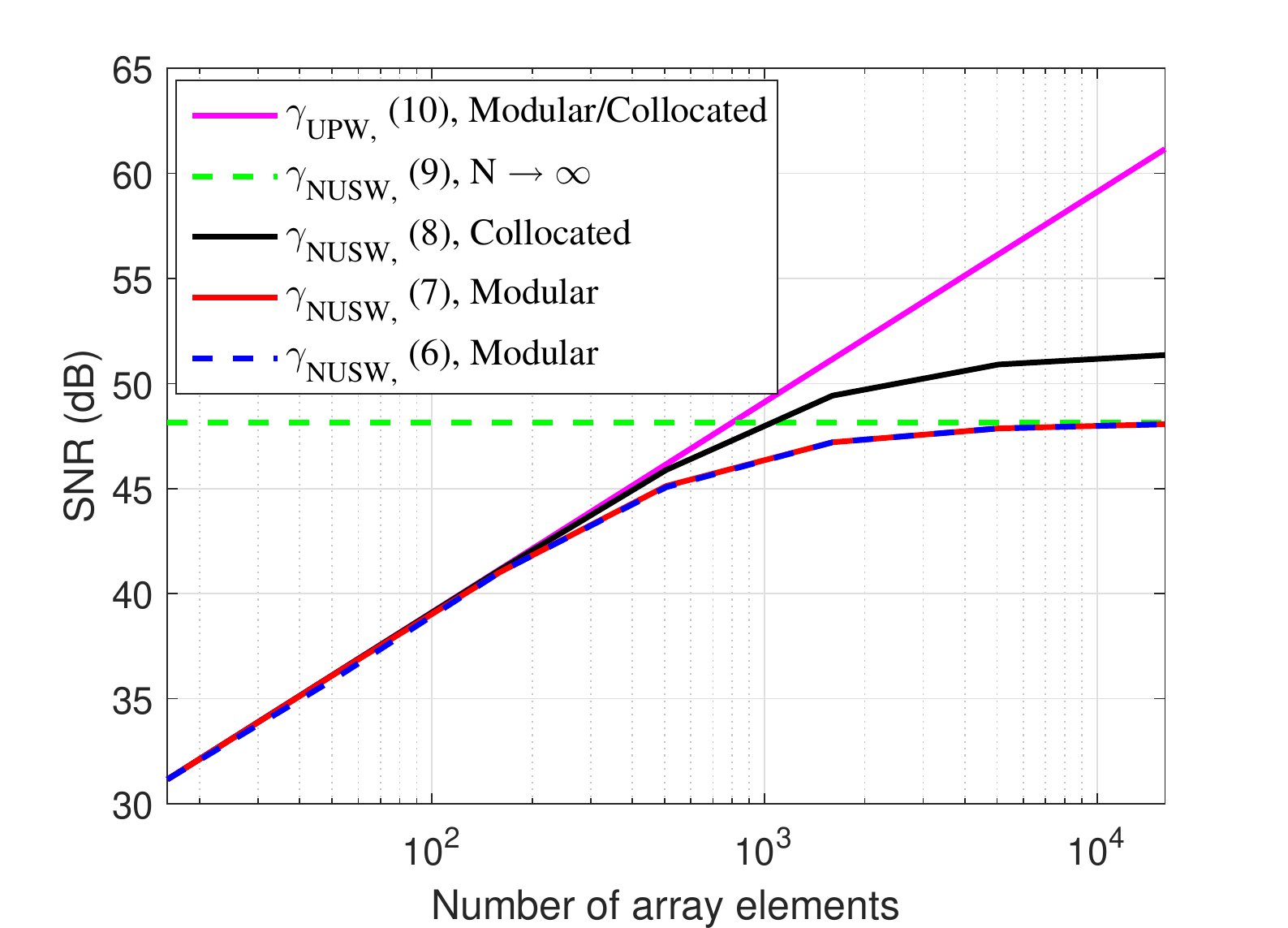}
  \vspace{-0.3cm}
  \caption{SNRs versus the number of array elements for different models.} \label{picture3}
  \end{centering}
  \vspace{-0.6cm}
  \end{figure}\par

  \begin{figure}[t]
    \begin{centering}
    \includegraphics[scale=0.45]{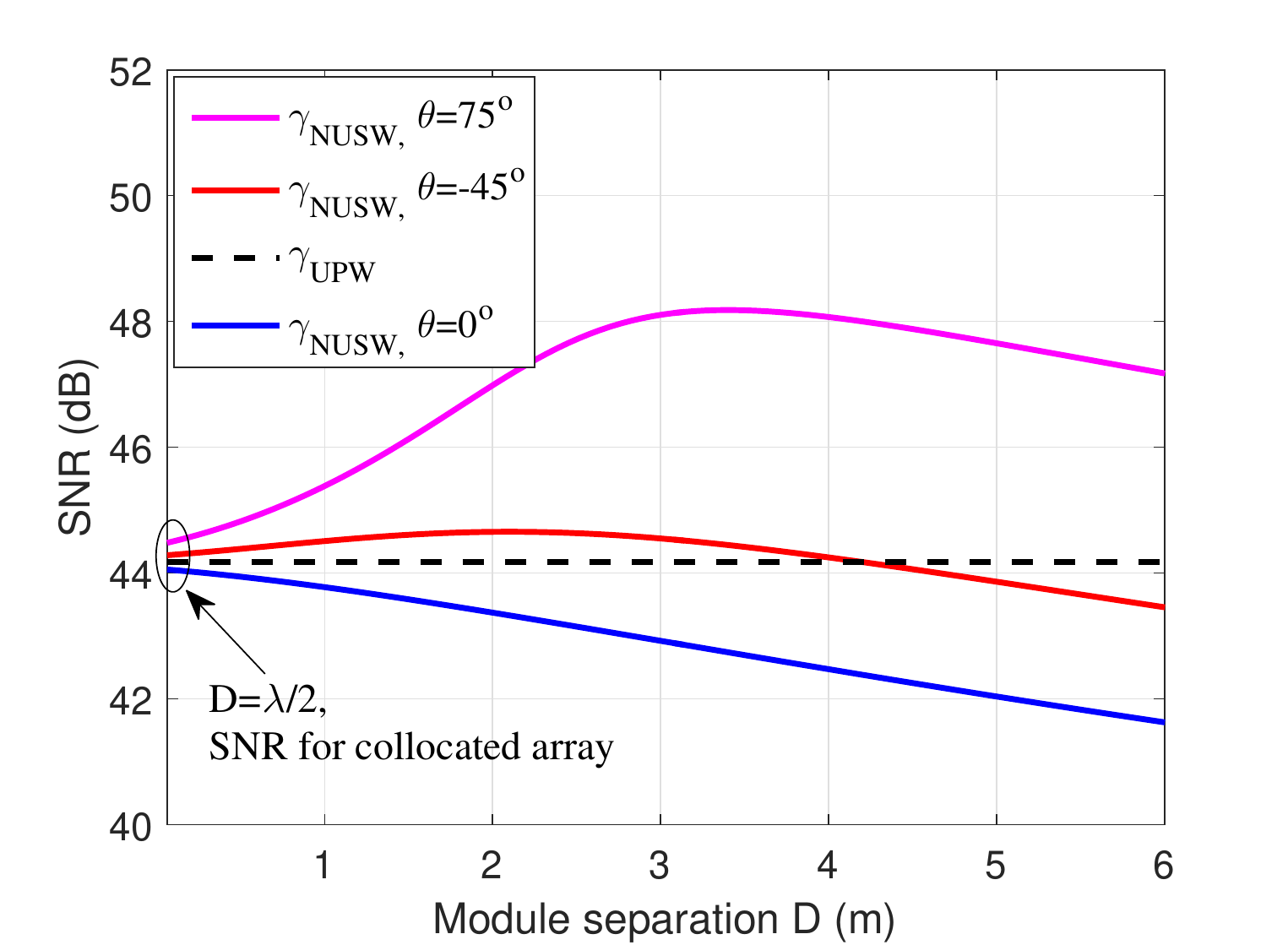}
    \vspace{-0.3cm}
    \caption{SNRs versus module separation $D$ with different user direction $\theta$.} \label{picture4}
    \end{centering}
    \vspace{-0.5cm}
    \end{figure}\par

In this section, numerical results are provided to
evaluate our developed results for modular XL-ULA communications.
The number of array elements in each module is $M=16$.
The carrier frequency is $2.4$ GHz, and inter-element spacing is $d=\frac{\lambda}{2}=0.0628$ $\rm m$.
The transmit SNR is $\overline P \beta_0=50$ $\rm dB$.
Unless otherwise specified,
the location of the user is ${\bf q}=[35,0]^T$ $\rm m $,
the module number is $N=20$,
and the module separation is $D=20d=1.256$ $\rm m$. \par
For modular array, Fig.~\ref{picture3} shows the SNRs $\gamma_{\rm NUSW}$ and $\gamma_{\rm UPW}$ versus the element number $MN$ by increasing $N$.
It is observed from Fig.~\ref{picture3} that
the closed-formed SNR expression in \eqref{EQU-6} perfectly matches with the summation expression in \eqref{EQU-5},
which verifies Theorem \ref{theo1}. For relatively small element number $MN$,
both $\gamma_{\rm NUSW}$ and $\gamma_{\rm UPW}$ increase linearly with $MN$, which is consistent with Corollary \ref{lemma3}.
However, as $MN$ further increases,
the results predicted by the NUSW and UPW models exhibit two different scaling laws, i.e.,
approaching to a constant versus growing unbounded.
This demonstrates the importance of properly modelling the spherical wave in modular XL-ULA communication.
In addition, Fig.~\ref{picture3} also plots the SNR with the conventional collocated array architecture.
It is observed that for the considered setup,
the difference of $\gamma_{\rm NUSW}$ for collocated XL-array and modular XL-array
is about $3$ dB for extremely large arrays.
This is expected since the modular XL-array
has extra inter-module separation for more flexible deployment.\par

Fig.~\ref{picture4} plots the SNRs $\gamma_{\rm NUSW}$ and $\gamma_{\rm UPW}$
versus the module separation $D$ with different user direction $\theta$,
where $D$ starts from $\frac{\lambda}{2}$ corresponding to the collocated array architecture.
As shown from Fig.~\ref{picture4}, $\gamma_{\rm UPW}$ is independent of $D$ and $\theta$, as expected.
By contrast, the SNR based on the NUSW model exhibits different trends for different user direction $\theta$.
It is also observed that depending on $\theta$, $\gamma_{\rm UPW}$ may either over-estimate (as for $\theta=0^o$) or under-estimate (as for $\theta=75^o$) the result predicted by the NUSW model. This demonstrates the necessity of NUSW modelling for modular XL-ULA communication.\par
\section{Conclusions}\label{g} \vspace{-1pt}
In this letter, we studied a novel modular XL-array architecture
to accommodate extremely large number of antennas while catering to the practical constraint of mounting structure.
The mathematical modelling and performance analysis for the modular XL-ULA were conducted,
by taking into account the NUSW characteristics.
The closed-form SNR expression was derived in terms of the total array size,
module separation, and the user's location.
Besides, the asymptotic SNR scaling law was revealed as the the array size enlarges by increasing the
 number of array modules.
Numerical results were provided to show the importance of near-field modelling for
characterizing the performance of modular XL-array communications.

\appendices

\section{Proof of Theorem 1}

We consider the case when $\theta \ne \pm \frac{\pi}{2}$ at first and define the function
$f(x,y)=\frac{1}{1-2x\sin\theta -2Ky\sin\theta +2Kxy +x^2+K^2y^2}$,
which is a continuous function over the domain $\mathcal{A}=\left\{(x,y)|-\frac{M}{2}\epsilon\le x \le\frac{M}{2}\epsilon, -\frac{N}{2}\epsilon\le y \le\frac{N}{2}\epsilon\right\}$.
Since $\epsilon \ll 1$, we have $f(x,y)\approx f(m\epsilon, n\epsilon)$, $\forall x \in [(m-\frac{1}{2})\epsilon, (m+\frac{1}{2})\epsilon]$ and $\forall y \in [(n-\frac{1}{2})\epsilon, (n+\frac{1}{2})\epsilon]$. Based on the concept of double integral, we have
\begin{equation}\label{EQU-12} \vspace{-3pt}
\begin{split}
\sum_{n=-\frac{N-1}{2}}^{\frac{N-1}{2}}\sum_{m=-\frac{M-1}{2}}^{\frac{M-1}{2}}f(m\epsilon, n\epsilon)\epsilon^2
\approx 	\int_{-\frac{N}{2}\epsilon}^{\frac{N}{2}\epsilon}\int_{-\frac{M}{2}\epsilon}^{\frac{M}{2}\epsilon} f(x,y) \, dx\,dy.\\
\end{split}
\end{equation}\par
By substituting $f(x,y)$ into \eqref{EQU-12},
we have \eqref{EQU-13}, shown at the top of the next page,
where $(a)$ follows from the integral formula 2.103 in \cite{Gradshteyn2007}, i.e., $\int \frac{1}{A+2Bx+Cx^2} \,dx=\frac{1}{\sqrt{AC-B^2}}\arctan{\frac{Cx+B}{\sqrt{AC-B^2}}}$, $AC>B^2$,
and $(b)$ always holds due to the integral formula 2.821 in \cite{Gradshteyn2007}, i.e., $\int \arctan \frac{x}{a} \,dx=x\arctan \frac{x}{a} -\frac{a}{2}\ln(a^2+x^2)$.
By substituting \eqref{EQU-13} into \eqref{EQU-5}, we have \eqref{EQU-6}.
Note that for special cases when $\theta= \pm \frac{\pi}{2}$ and $r>\frac{S_1}{2}$, \eqref{EQU-6} can be verified easily by taking its limit. The proof of Theorem 1 is therefore completed.
\begin{figure*}[ht]
	\begin{equation}\label{EQU-13}
    \begin{split}	&\int_{-\frac{N}{2}\epsilon}^{\frac{N}{2}\epsilon}\int_{-\frac{M}{2}\epsilon}^{\frac{M}{2}\epsilon} \frac{1}{1-2x\sin\theta -2K y\sin\theta +2Kxy+x^2+K^2y^2} \, dx\,dy\\
    &\overset{(a)}{=}\frac{1}{\cos\theta}\int_{-\frac{N}{2}\epsilon}^{\frac{N}{2}\epsilon} \arctan\left(\frac{2Ky+M\epsilon-2\sin\theta}{2\cos\theta}\right) \,dy
    -\frac{1}{\cos\theta}\int_{-\frac{N}{2}\epsilon}^{\frac{N}{2}\epsilon} \arctan\left(\frac{2Ky-M\epsilon-2\sin\theta}{2\cos\theta}\right)\,dy\\
    &\overset{(b)}{=}\frac{1}{K}\left[h\left(\frac{S_1}{2r\cos\theta}-\tan\theta\right)+h\left(\frac{S_1}{2r\cos\theta}+\tan\theta\right)
    -h\left(\frac{S_1-2Md}{2r\cos\theta}-\tan\theta\right)-h\left(\frac{S_1-2Md}{2r\cos\theta}+\tan\theta\right)\right].\\
    \end{split}
	\end{equation}
{\noindent} \rule[-10pt]{17.5cm}{0.05em}
\end{figure*}

\section{Proof of Corollary 1}
We first define $g(x)\overset{\Delta}{=}h(Mx)=Mx\arctan(Mx)-\frac{1}{2}\ln(1+M^2x^2)$,
and thus \eqref{EQU-6} can be rewritten as
\begin{equation}\label{EQU-714} \vspace{-3pt}
\begin{split}
&\gamma_{\rm NUSW}\approx\\
&\frac{{\overline P}\beta_0}{M d^2}\left[g\left(\frac{(N+1)d}{2r\cos\theta}-\frac{\tan\theta}{M}\right)+g\left(\frac{(N+1)d}{2r\cos\theta}+\frac{\tan\theta}{M}\right)\right.\\
&\left.-g\left(\frac{(N-1)d}{2r\cos\theta}-\frac{\tan\theta}{M}\right)-g\left(\frac{(N-1)d}{2r\cos\theta}+\frac{\tan\theta}{M}\right)\right].\\
\end{split}
\end{equation}\par
Then, by using first-order Taylor series expansion of $g\left(\frac{(N+1)d}{2r\cos\theta}-\frac{\tan\theta}{M}\right)$ for
$x_0=\frac{Nd}{2r\cos\theta}-\frac{\tan\theta}{M}$, we have
$g\left(\frac{(N+1)d}{2r\cos\theta}-\frac{\tan\theta}{M}\right)\approx  g\left(\frac{Nd}{2r\cos\theta}-\frac{\tan\theta}{M}\right)+\frac{Md}{2r\cos\theta}\arctan\left(\frac{MNd}{2r\cos\theta}-\tan\theta\right)$.
Similarly, we can obtain the approximations for other three terms inside the bracket of \eqref{EQU-714}.
Lastly, by substituting these approximation results into \eqref{EQU-714}, we have \eqref{EQU-7151}. The proof of Corollary \ref{lemma1} is completed.\par

\section{Proof of Corollary 2}
When $N \to \infty$, \eqref{EQU-6} can be expressed as the form of the sum of two limit terms, i.e.,
\begin{equation}\label{EQU-17} \vspace{-3pt}
\begin{split}
&\lim_{N \to \infty}\gamma_{\rm NUSW}= \\
&\frac{{\overline P}\beta_0}{K d^2}\left\{\lim_{N \to \infty}\left[h\left(\frac{S_1}{2r\cos\theta}-\tan\theta\right)-h\left(\frac{S_1-2Md}{2r\cos\theta}-\tan\theta\right)\right]\right.\\
&\left.+\lim_{N \to \infty}\left[h\left(\frac{S_1}{2r\cos\theta}+\tan\theta\right)-h\left(\frac{S_1-2Md}{2r\cos\theta}+\tan\theta\right)\right]\right\}.\\
\end{split}
\end{equation}\par
For the first limit term, we have

\begin{equation}\label{EQU-151} \vspace{-3pt}
\begin{split}
&\lim_{N \to \infty}\left[h\left(\frac{S_1}{2r\cos\theta}-\tan\theta\right)-h\left(\frac{S_1-2Md}{2r\cos\theta}-\tan\theta\right)\right]=\\
&\lim_{N \to \infty}\left\{\left(\frac{Kd}{2r\cos\theta}N-\frac{Md+2r\sin\theta}{2r\cos\theta}\right)\left[\arctan\left(\frac{Kd}{2r\cos\theta}N\right.\right.\right.\\
&\left.+\frac{Md-2r\sin\theta}{2r\cos\theta}\right)-\arctan\left(\frac{Kd}{2r\cos\theta}N\left.\left.\left.-\frac{Md+2r\sin\theta}{2r\cos\theta}\right)\right]\right\}\right.\\
&+\lim_{N \to \infty}\left[\frac{Md}{r\cos\theta}\arctan\left(\frac{Kd}{2r\cos\theta}N+
\frac{Md-2r\sin\theta}{2r\cos\theta}\right)\right]\\
&-\lim_{N \to \infty}\frac{1}{2}\ln\left(\frac{1+\left(\frac{Kd}{2r\cos\theta}N+\frac{Md-2r\sin\theta}
{2r\cos\theta}\right)^2}{1+\left(\frac{Kd}{2r\cos\theta}N-\frac{Md+2r\sin\theta}{2r\cos\theta}\right)^2}\right)\\
&\overset{(a)}{=}\frac{\pi Md}{2r\cos\theta},\\
\end{split}
\end{equation}
where $(a)$ holds due to hospital's rule, and the limit formulas include $\lim_{x \to  +\infty}\arctan x=\frac{\pi}{2}$
and $\lim_{x\to +\infty}\frac{a_1x^2+b_1x+c_1}{a_2x^2+b_2x+c_2}=\frac{a_1}{a_2}$.
Similarly, it is not difficult to obtain that
$\lim_{N \to \infty}\left[h\left(\frac{S_1}{2r\cos\theta}+\tan\theta\right)-h\left(\frac{S_1-2Md}{2r\cos\theta}+\tan\theta\right)\right]
=\frac{\pi Md}{2r\cos\theta}$.
By substituting the above two limit values into \eqref{EQU-17}, we have \eqref{EQU-7}.
The proof of Corollary \ref{lemma2} is thus completed.\par

\section{Proof of Corollary 3}
Similar to the proof of Corollary 1, by letting $g(x)=h(Mx)$, \eqref{EQU-6} can be expressed as
\begin{equation}\label{EQU-7141} \vspace{-3pt}
\begin{split}
&\gamma_{\rm NUSW}\approx \frac{{\overline P}\beta_0}{K d^2}\left[g\left(\frac{(\frac{K}{M}N+1)d}{2r\cos\theta}-\frac{\tan\theta}{M}\right)\right.\\
&+g\left(\frac{(\frac{K}{M}N+1)d}{2r\cos\theta}+\frac{\tan\theta}{M}\right)-g\left(\frac{(\frac{K}{M}N-1)d}{2r\cos\theta}-\frac{\tan\theta}{M}\right)\\
&\left.-g\left(\frac{(\frac{K}{M}-1)d}{2r\cos\theta}+\frac{\tan\theta}{M}\right)\right].\\
\end{split}
\end{equation}\par
 Based on the first-order Taylor series expansion of $g\left(\frac{(\frac{K}{M}N+1)d}{2r\cos\theta}-\frac{\tan\theta}{M}\right)$
 for $x_0=\frac{\frac{K}{M}Nd}{2r\cos\theta}-\frac{\tan\theta}{M}$,
 we have
 $g\left(\frac{(\frac{K}{M}N+1)d}{2r\cos\theta}-\frac{\tan\theta}{M}\right)\approx g\left(\frac{\frac{K}{M}Nd}{2r\cos\theta}-\frac{\tan\theta}{M}\right)+\frac{Md}{2r\cos\theta}\arctan\left(\frac{KNd}{2r\cos\theta}-\tan\theta\right)$.

By following the similar procedure, the first-order Taylor approximations for other three terms inside the bracket of \eqref{EQU-7141}
can be obtained.\par
By combining the above results, \eqref{EQU-7141} is given by
\begin{equation}\label{EQU-719} \vspace{-3pt}
\gamma_{\rm NUSW}\approx\frac{M{\overline P}\beta_0}{K d r \cos\theta}\Gamma,\\
\end{equation}
where $\Gamma=\arctan\left(\frac{KNd}{2r\cos\theta}-\tan\theta\right)+\arctan\left(\frac{KNd}{2r\cos\theta}+\tan\theta\right)$.
Based on the geometric equivalence in \cite{Lu2021}, when $r \gg \frac{1}{2}S_1$, we have
\begin{equation}\label{EQU-71112} \vspace{-3pt}
\begin{split}
\Gamma&=\arctan\left(\frac{\frac{KNd}{2}\cos\theta}{r-\frac{KNd}{2}\sin\theta}\right)+\arctan\left(\frac{\frac{KNd}{2}\cos\theta}{r+\frac{KNd}{2}\sin\theta}\right)\\
&\approx 2\arctan\left(\frac{KNd}{2r}\cos\theta\right)\overset{(a)}{\approx} \frac{KNd}{r}\cos\theta,\\
\end{split}
\end{equation}
where $(a)$ holds due to the fact that $\arctan x \approx x$ for $|x| \ll 1$.
By substituting \eqref{EQU-71112} into \eqref{EQU-719}, we have \eqref{EQU-11}, and Corollary \ref{lemma3} is proved.


\end{document}